\documentclass[prl,superscriptaddress,towcolumn,reprint]{revtex4-2}

\usepackage{graphicx}
\usepackage{dcolumn}
\usepackage{bm}
\usepackage{color}
\usepackage{amsmath,amsfonts,amssymb}
\usepackage{graphicx}
\usepackage[english]{babel}
\usepackage{color}
\usepackage{booktabs,longtable}
\usepackage{natbib}
\usepackage{textgreek} 
\usepackage{upgreek} 
\usepackage{mathrsfs}
\usepackage{array}
\usepackage{booktabs,tabularx, colortbl}
\usepackage{multirow}
\usepackage{lipsum}

\usepackage[colorlinks]{hyperref}
\hypersetup{
                pdfstartview={FitH},
                linkcolor=blue,
                citecolor=blue,
                filecolor=blue,
                urlcolor=blue
}

\newcommand\varpm{\mathbin{\vcenter{\hbox{
  \oalign{\hfil$\scriptstyle+$\hfil\cr
          \noalign{\kern-.3ex}
          $\scriptscriptstyle({-})$\cr}
}}}}
\newcommand\varmp{\mathbin{\vcenter{\hbox{
  \oalign{\hfil$\scriptscriptstyle-$\hfil\cr
          \noalign{\kern-.3ex}
          $\scriptstyle({+})$\cr}
}}}}

\newcolumntype{Y}{>{\centering\arraybackslash}X}

\begin{document}

\begin{sloppypar}

\title{Electric Field Modulation of Spin Transport}


\author{C.~Zucchetti}
\email{carlo.zucchetti@polimi.it\\}
\affiliation{LNESS-Dipartimento di Fisica, Politecnico di Milano, Piazza Leonardo da Vinci 32, 20133 Milan, Italy}
\author{A.~Marchionni}
\affiliation{LNESS-Dipartimento di Fisica, Politecnico di Milano, Piazza Leonardo da Vinci 32, 20133 Milan, Italy}
\author{M.~Bollani}
\affiliation{Istituto di Fotonica e Nanotecnologie del Consiglio Nazionale delle Ricerche, Piazza Leonardo da Vinci 32, 20133 Milan, Italy}
\author{F.~Ciccacci}
\affiliation{LNESS-Dipartimento di Fisica, Politecnico di Milano, Piazza Leonardo da Vinci 32, 20133 Milan, Italy}
\author{M.~Finazzi}
\affiliation{LNESS-Dipartimento di Fisica, Politecnico di Milano, Piazza Leonardo da Vinci 32, 20133 Milan, Italy}
\author{F.~Bottegoni}
\affiliation{LNESS-Dipartimento di Fisica, Politecnico di Milano, Piazza Leonardo da Vinci 32, 20133 Milan, Italy}


\date{\today}


\begin{abstract}

The finite spin lifetime in solids is often considered a major hindrance for the development of spintronic devices, which typically require cryogenic temperatures to mitigate this phenomenon. In this work we show that this feature can instead be exploited to realize a scheme where spin transport is modulated at room temperature by a modest electric field. A field directed antiparallel (parallel) to the spin-diffusion velocity can in fact largely increase (decrease) the spin-transport length compared with the zero field case. We find that applying an electric field ${E=24~\text{V/cm}}$ along a $40~\text{\textmu m}$-long path in germanium results in about one order of magnitude modulation of the spin-polarized electrons entering in the detector. The present work demonstrates that electric fields can be exploited for guiding spins over macroscopic distances and for realizing fast, room temperature modulation of spin accumulation. 

\end{abstract}


\maketitle
\medskip


The ultimate goal of spintronics is the active control of the spin-polarization within a solid-state environment \cite{Awschalom2007,Fabian2007}. Recently, spin-orbit coupling (SOC) emerged as a promising tool for implementing such a spin manipulation \cite{Manchon2015}. A large SOC, however, leads to short spin lifetime and diffusion lengths, representing a severe limitation for the development of spintronic devices. Conversely, the small SOC in light semiconductors yields to longer electron spin lifetimes, as observed in both lightly-doped bulk \cite{Hamaya2018} and low-dimensional materials \cite{Watson2017,Dirnberger2019}. Among semiconductors, Si and Ge possess the longest lifetimes, since, at variance from GaAs, the Dyakonov-Perel mechanism of spin relaxation is totally suppressed in a centrosymmetric lattice \cite{Dyakonov1972}. Despite Si-based spintronics would be desirable, generating non-equilibrium spin populations in lightly-doped Si is inefficient compared to Ge \cite{Rioux2010,Cheng2011}. Moreover, Ge-based platforms could host spintronics, electronics, and photonics devices, all integrated on a Si-compatible platform. Indeed, the Ge direct gap matches the conventional telecommunication window \cite{Hochberg2010,Suess2013}, and the $4\%$ lattice mismatch with Si does not prevent the ready integration of Ge with the mainstream Si-based technology, eventually enabling strain to engineer new functionalities \cite{Bottegoni2011,Li2012a,Bollani2015}.

Within this frame, a relevant building block of semiconductor spintronic would be to carry the spin information over long distances. However, spin transport in semiconductors have been mostly explored in the diffusive regime. While the application of electric fields in ferromagnet/semiconductor structures has been demonstrated to increase the spin injection into semiconductors by orders of magnitude \cite{Yu2002a,Hanbicki2002}, only few experimental studies focus on gate control of spin transport in semiconductors \cite{Kwon2008,Wang2013}. Moreover, despite a variety of theoretical studies describe the drift-diffusive spin transport \cite{Yu2002a,Miah2008}, a clear experimental proof of this regime is still lacking.

We have recently proposed a nonlocal spin-injection/detection scheme by which we have been able to directly image spin transport in diffusive regime \cite{Zucchetti2018a,Bottegoni2020}. We now employ the same paradigm to investigate drift-diffusive spin transport, showing that a finite spin-diffusion length combined with the application of an electric field can be capitalized to manipulate spin accumulation. The investigated scheme could in principle yield modulation of the spin transport both on long distances and within timescales faster than other spin accumulation/modulation systems.

In particular, we report on the drift-diffusive spin transport regime in lightly $n$-doped Ge at room temperature. We exploit the optical orientation technique \cite{Lampel1968,Meier1984} to locally generate spin-polarized electrons in Ge. The spin detection is performed by measuring the voltage drop given by the inverse spin-Hall effect (ISHE) in a thin Pt pad \cite{Dyakonov1971a,Dyakonov1971b} grown on Ge and spatially separated from the point where spin is generated. This nonlocal spin injection/detection scheme allows one to directly image spin transport either in the diffusive \cite{Zucchetti2017a} and in the drift-diffusive regime. We find that the typical decay length of the spin polarization $L_s$ is doubled (halved) compared with the diffusive case for spin-polarized electrons flowing downstream (upstream) with respect to the applied electric field ${E=24~\text{V/cm}}$. In the best case scenario, we experimentally detect a spin loss of just the $40\%$ within a ${30~\text{\textmu m}}$-long path, corresponding to ${L_{s}\approx40~\text{\textmu m}}$. Comparable values of the spin-diffusion length in semiconductors have been predicted \cite{Cheng2010,Li2011,Song2012,Li2012a,Finazzi2016} and observed \cite{Bottegoni2017b} only at cryogenic temperatures.

\begin{figure*}[t]
\begin{center}
\includegraphics[width=0.85\textwidth]{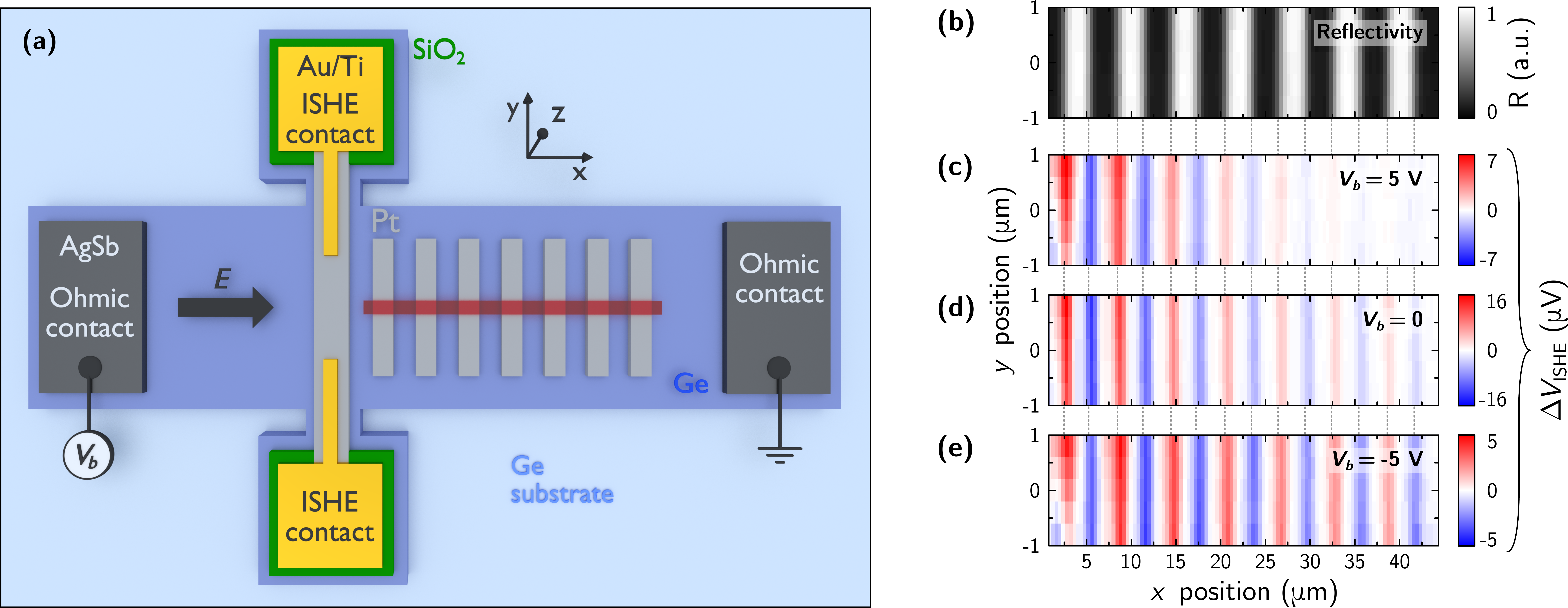}
\caption{(a) Sketch of the sample. A mesa of Ge is realized on the top of a Ge substrate. A set of Pt stripes is employed for spin injection, while a contacted Pt stripe works as a spin detector by means of ISHE. AgSb Ohmic contacts are exploited to generate an electrostatic field $E$ in Ge. (b-e) Experimental data showing (b) the refelctivity map of the acquired region [highlighted in red in panel (a)] and the corresponding $\Delta V_{\text{ISHE}}$ signal for a bias voltage equal to (c) ${V_b=5~\text{V}}$, (d) ${V_{b}=0}$ and (e) ${V_{b}=-5~\text{V}}$.}\label{fig:1}
\end{center}
\end{figure*}

A sketch of the investigated sample is reported in Fig.~\ref{fig:1}\textcolor{blue}{(a)}. We etch a lightly $n$-doped Ge substrate (donor concentration ${N_d\approx2\times10^{15}~\text{cm}^{-3}}$) to obtain a ${1~\text{\textmu m}}$-thick and ${50~\text{\textmu m}\times1~\text{mm}}$-wide mesa, to loosely constrain the spin transport along one direction. We then lithographically define a set of Pt stripes on the mesa surface, which are employed for spin-injection (see below). These stripes are ${10~\text{nm}}$-thick and ${40\times3~\text{\textmu m}^2}$-wide, with a center-to-center pitch of ${6~\text{\textmu m}}$. One of them [the first and ${5~\text{\textmu m}}$-wide stripe on the left in Fig.~\ref{fig:1}\textcolor{blue}{(a)}] is contacted with Au/Ti electrodes and serves as a spin detector. The flow of spin-polarized electrons entering this Pt detection pad is converted via the inverse spin-Hall effect (ISHE) in a charge flow, which we record as a voltage drop in open-circuit conditions across the ISHE electrodes. A $\text{SiO}_2$ layer [green in Fig.~\ref{fig:1}\textcolor{blue}{(a)}] below the contacts on the ISHE detector prevents direct spin and charge absorption from the substrate. We apply a bias electric potential $V_b$ across two AgSb Ohmic contacts \cite{Dumas2014} grown on Ge to generate a static electric field $E$ in the semiconductor. For further details on sample growth and characterization see the Supplemental Material.

The measurements are performed at room temperature with a confocal microscope (see Ref.~\onlinecite{Zucchetti2017a}). The light source is a laser diode tuned at the direct gap of Ge (${h\nu=0.8~\text{eV}}$) delivering an optical power equal to ${270~\text{\textmu W}}$. The circular polarization of the photons is modulated at ${50~\text{kHz}}$ by means of a photoelastic modulator (PEM). An objective with $0.7$ numerical aperture focuses the light on the sample with a diffraction-limited spot of ${\approx1.3~\text{\textmu m}}$. The measured ISHE signal ${\Delta V_{\text{ISHE}}}$ is obtained by demodulating at the PEM frequency the voltage drop recorded across the Au/Ti ISHE contacts, while the laser spot raster scans the surface of the sample. The optical reflectivity of the sample is simultaneously measured with a photodetector.

In our setup, solely the electrons with a component of the spin polarization directed along $x$ [reference frame of Fig.~\ref{fig:1}\textcolor{blue}{(a)}] give rise to a detectable $\Delta V_{\text{ISHE}}$ (as detailed, e.g., in Ref.~\onlinecite{Zucchetti2018a}). When the light beam impinges on the edge of a metal stripe, the electromagnetic wavefront is altered \cite{Bottegoni2014} yielding a highly-localized spin accumulation in the semiconductor, with a detectable in-plane positive (negative) component along the $x$ axis below the left (right) edge of the Pt stripe. The details of the technique are reported in Refs.~\citenum{Bottegoni2014,Zucchetti2017a,Guillet2020b}. 

The spin-polarized electrons are generated at the $\Gamma$ point of the Brillouin zone and within $300~\text{fs}$ they scatter to the $L$ valleys, mostly preserving their spin polarization \cite{Pezzoli2012,Zucchetti2018a}. Within the explored range of applied electric fields (${\left|\mathbf{E}\right|<24~\text{V/cm}}$) conduction occurs at $L$ \cite{Gunn1964,Okamoto2014}. It is worth mentioning that, together with spin-polarized electrons, a net population of spin-polarized holes is also generated via optical orientation \cite{Rioux2010}. However, any hole contribution to the spin current over the investigated length scales can be neglected since the hole spin lifetime is only a few picoseconds \cite{Rortais2016a}. This limits the spin-diffusion length of holes to few tens of nanometers in the purely diffuive regime and to --~possibly~-- a hundred nanometers in the drift-diffusion regime. Note that the minimum distance between generation and detection points in our setup is ${3~\text{\textmu m}}$. 

In Figs.~\ref{fig:1}\textcolor{blue}{(b-e)} we report the experimental map acquired in the ${40\times2~\text{\textmu m}^2}$ band highlighted in red in Fig.~\ref{fig:1}\textcolor{blue}{(a)}. Due to the higher reflectivity of Pt stripes compared to the Ge substrate, the former appear bright in the reflectivity map [Fig.~\ref{fig:1}\textcolor{blue}{(b)}]. Panels \textcolor{blue}{(c-e)} show the map of the electrical signal $\Delta V_{\text{ISHE}}$ recorded by the detector (located out of the maps at ${x=0}$) for three values of the bias voltage $V_{b}$. In all cases, when the light beam illuminates the left (right) edges of Pt stripes a positive (negative) $\Delta V_{\text{ISHE}}$ is detected, corresponding to the complementary spin populations generated in Ge below opposite Pt edges. This, together with the linear dependence of $\Delta V_{\text{ISHE}}$ on the optical power and on the degree of circular polarization of the light (data not shown) confirms the spin-related nature of the detected signal \cite{Zucchetti2018a}.

\begin{figure}[t]
\begin{center}
\includegraphics[width=0.48\textwidth]{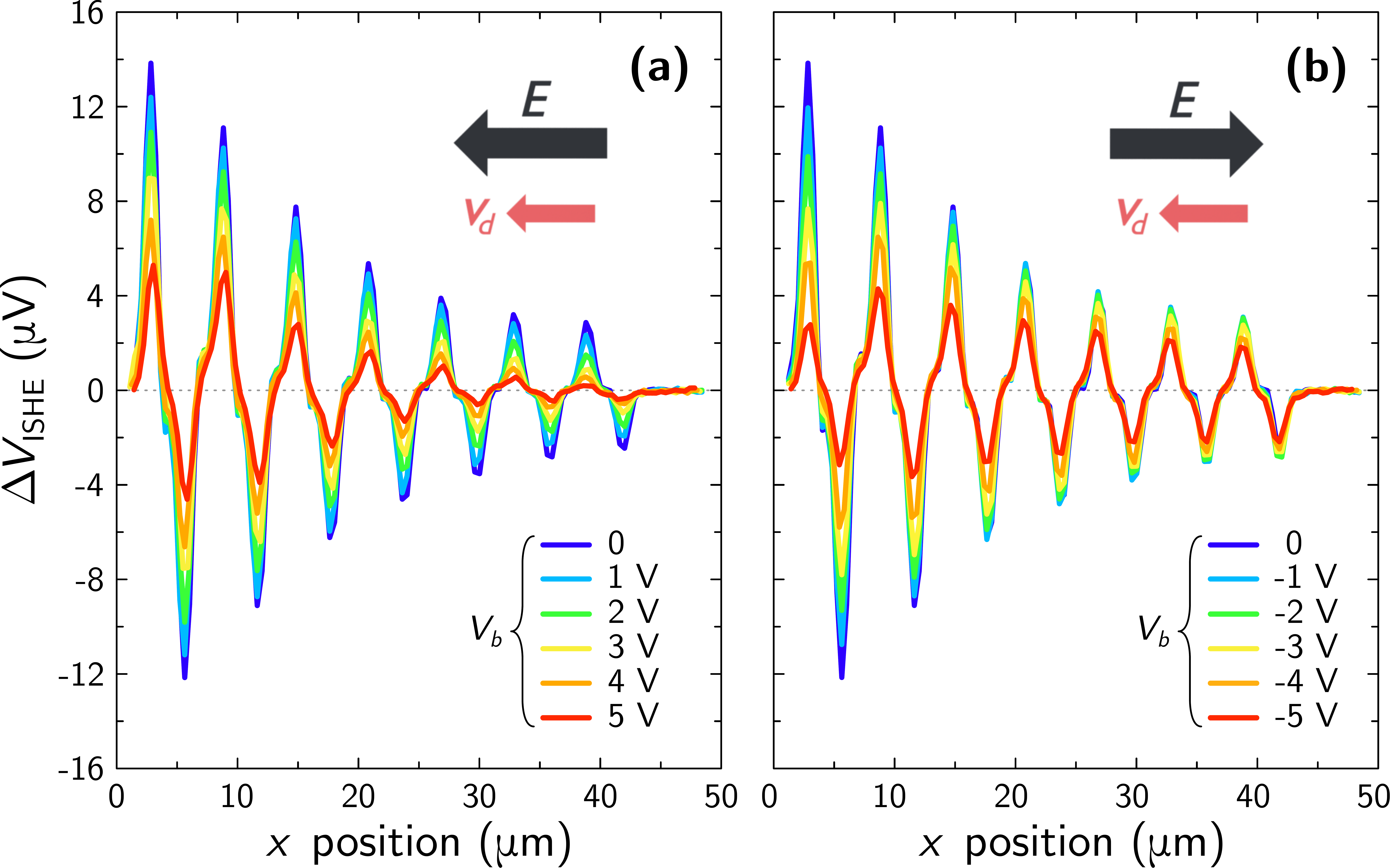}
\caption{Profiles of the ISHE signal acquired in the red band shown in Fig.~\ref{fig:1}\textcolor{blue}{(a)} and averaged along the $y$ axis for (a) positive and (b) negative $V_b$ values. $E$ and $v_d$ arrows represent the direction of the electric field and of the spin-diffusion velocity. Note that all the experimental data are acquired with the same optical power, equal to ${270~\text{\textmu W}}$.}\label{fig:2}
\end{center}
\end{figure}

In Fig.~\ref{fig:1}\textcolor{blue}{(c-e)}, the absolute value of the ISHE signal decreases with increasing the distance of the injection point of the spin-polarized electrons, i.e., at larger $x$ values. This is the fingerprint of the spin depolarization that electrons experience before reaching the detector. Such a decrease is less (more) pronounced for negative (positive) $V_b$ values, corresponding to ${E_x>0}$ (${E_x<0}$) [see panels \textcolor{blue}{(e)} and \textcolor{blue}{(c)}, respectively] compared to the zero field case [panel \textcolor{blue}{(d)}]. This indicates that the transport of spin-polarized electrons can be modulated by the application of an electric field: depending on the field direction, electrons acquire a drift velocity and cover the distance to the ISHE detector in less (more) time, thus reducing (increasing) their loss of spin polarization. 
It is worth mentioning that, despite high electric fields could contribute to a faster spin depolarization, the employed $E$ values are not expected to significantly affect the spin-relaxation time \cite{Yu2002a,Li2012}. 

In Fig.~\ref{fig:2} we show the average along the $y$ axis of the $\Delta V_{\text{ISHE}}$ maps similar to those reported in Fig.~\ref{fig:1}\textcolor{blue}{(c-e)}, for a set of positive [panel \textcolor{blue}{(a)}] and negative [panel \textcolor{blue}{(b)}] bias voltages. Due to the modulation of the typical decay length of the spin-polarization  --~spin-transport length $L_s$, from now on~-- spin-polarized electrons generated far from the detection point are suppressed (preserved) as ${\left|V_b\right|}$ increases for positive (negative) values. In the most favorable scenario, corresponding to ${V_b=-5~\text{V}}$, a ${30~\text{\textmu m}}$-long path reduces the $\Delta V_{\text{ISHE}}$ signal by $40\%$ only. Moreover, $\Delta V_{\text{ISHE}}$ is modulated by about one order of magnitude when optical spin orientation is performed at the last Pt stripe ($x\approx40~\text{\textmu m}$) by switching the bias voltage between ${V_b=+5~\text{V}}$ and ${V_b=-5~\text{V}}$.

It can be noticed that in both Fig.~\ref{fig:1}\textcolor{blue}{(c-e)} and Fig.~\ref{fig:2} the maximum value of $\Delta V_{\text{ISHE}}$ is obtained for ${V_b=0}$. For ${V_b>0}$, the lowering of $\Delta V_{\text{ISHE}}$ is due to the reduction of the spin-transport length in the drift-diffusion regime. Conversely, for ${V_b<0}$, the applied field quickly removes electrons from below the detection pad, reducing their transit time and, consequently, the probability that spin-polarized electrons are injected into the detection pad and contribute to $\Delta V_{\text{ISHE}}$. 

\begin{figure}[b]
\begin{center}
\includegraphics[width=0.48\textwidth]{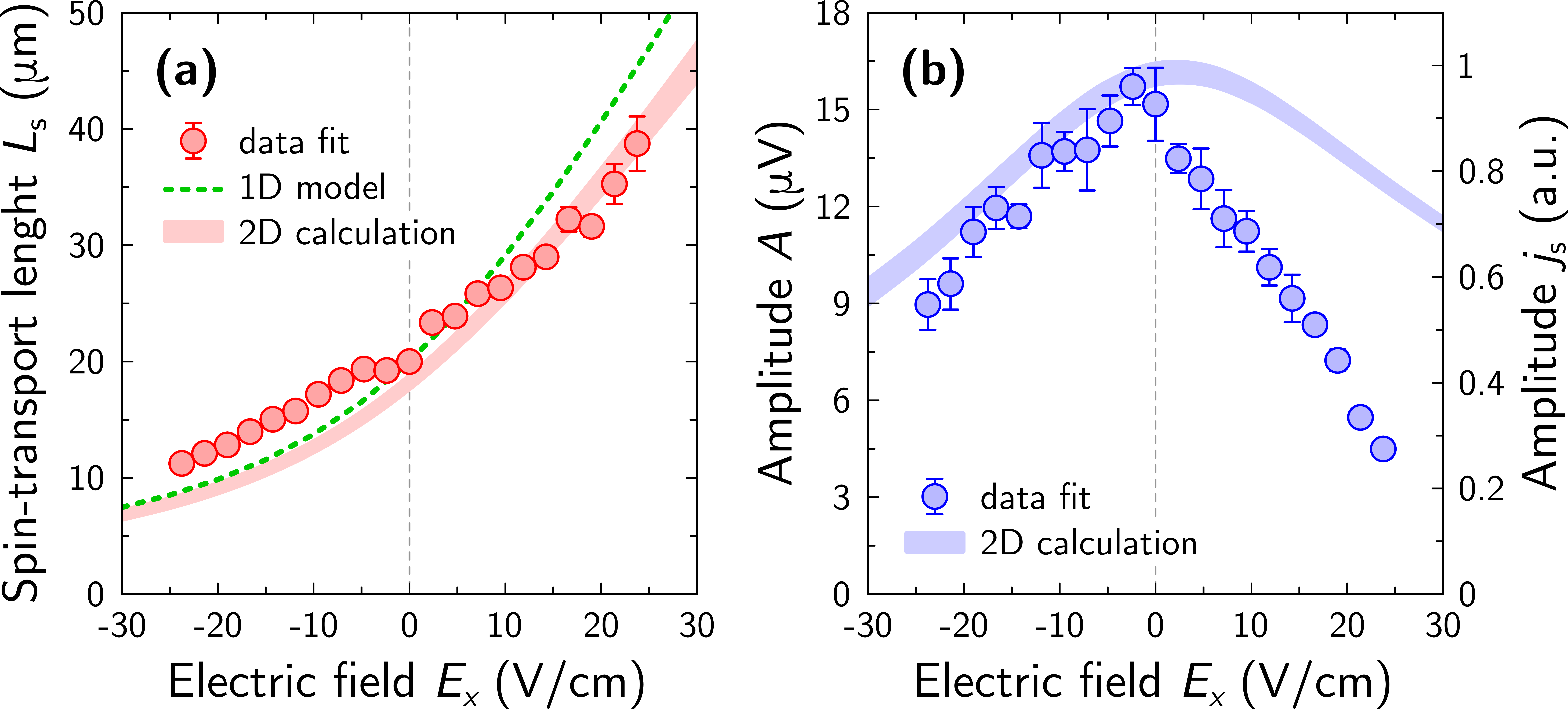}
\caption{(a) Spin-transport length $L_{\text{s}}$ and (b) amplitude related to spin-sink efficiency $A$ (see text) obtained from the fitting of experimental profiles similar to the ones reported in Fig.~\ref{fig:2}. The green dotted line in panel (a) represents the spin-transport length predicted by the analytical one-dimensional spin drift-diffusion model [Eq.~\eqref{eq:1}]. The thick lines represent the results of two-dimensional numerical simulations (see text).}\label{fig:3}
\end{center}
\end{figure}

We perform a quantitative analysis of the profiles in Fig.~\ref{fig:2} by fitting the absolute value of the $\Delta V_{\text{ISHE}}$ peaks with a function ${A\,e^{-x/L_s}}$ that accounts for spin depolarization in the generation-to-detection path, being $A$ the amplitude related to spin sink efficiency of the detector and $L_s$ the spin-transport length \cite{Zucchetti2017a}. For ${V_b=0}$, the latter corresponds to the spin-diffusion length ${L_{s,0}=\sqrt{D\tau_{s}}}$, being $D$ the diffusion coefficient and $\tau_s$ the spin lifetime. The $L_s$ and $A$ values obtained from the fitting procedure are reported in Fig.~\ref{fig:3} as a function of the applied electric field $E_x$. The latter has been estimated from the bias potential $V_b$ by numerically simulating the three-dimensional field distribution in the whole structure. We obtain that a bias potential ${V_b=1~\text{V}}$ yields an almost uniform electric field ${E_x=-4.75~\text{V/cm}}$ into the stripe. Further information about the electrical simulations are reported in the Supplemental Material. 

While the ${L_{s,0}=20\pm0.5~\text{\textmu m}}$ value obtained at ${E_x=0}$ is comparable with previous results obtained with lightly $n$-doped Ge \cite{Zucchetti2019b}, $L_{\text{s}}$ is nearly doubled (halved) compared to $L_{s,0}$ at ${E_x=\pm24~\text{V/cm}}$ for spin-polarized electrons flowing downstream (upstream) [see panel \textcolor{blue}{(a)}]. The largest value we obtain is ${L_s=39\pm2~\text{\textmu m}}$, a distance previously achieved only in cryogenic conditions \cite{Bottegoni2017b}. In Fig.~\ref{fig:3}\textcolor{blue}{(a)} the experimentally inferred $L_s$ values are compared to the spin-transport length obtained from the analytical solution of the one-dimensional (1D) coupled spin continuity and spin drift-diffusion equations (green dotted line, details in the Supplemental Material) \cite{Yu2002a,Miah2008}:
\begin{equation}\label{eq:1}
L_s=\left[-\frac{1}{2}\frac{eE_x}{k_{\text{B}}T}+\sqrt{\left(\frac{1}{2}\frac{eE_x}{k_{\text{B}}T}\right)^2+\frac{1}{L_{s,0}^2}}~\right]^{-1},
\end{equation}
being $k_{\text{B}}$ the Boltzmann constant, $e$ the elementary charge, and $T$ the lattice temperature. Eq.~(\ref{eq:1}), which nicely reproduces the experimental $L_s$ vs. $E_x$ trend reported in Fig.~\ref{fig:1}\textcolor{blue}{(a)}, reveals also the crucial role of the \textit{finite} spin diffusion length $L_{s,0}$ for the electrical modulation of spin transport. In fact, according to Eq.~(\ref{eq:1}), a significant electrical modulation of $L_s$ and, hence, of the spin accumulation along the device, is only possible for ${L_{s,0}\lesssim2 k_\text{B}T/(eE_x)}$.

To better understand the role of geometry in our device, we performed two-dimensional (2D) finite-element spin-transport simulations on the $xz$ plane in Fig.~\ref{fig:1}\textcolor{blue}{(a)}. To keep the numerical simulations as simple as possible, we assumed a uniform electric field directed along $x$ (details in the Supplemental Material). We then estimated the spin current density $j_s$ entering the Pt detector, which is proportional to the experimentally measured $\Delta V_{\text{ISHE}}$ value \cite{Ando2010}. Compared to the 1D model, the 2D simulations slightly improve the agreement with the experimentally inferred $L_{\text{s}}$ value at positive $E$, as shown by the thick line in Fig.~\ref{fig:3}\textcolor{blue}{(a)}. The difference between the theoretical and experimental values of $L_s$ is lower than $2~\text{\textmu m}$ within the whole explored range of applied electric fields.

The 2D simulations also qualitatively reproduce the trend of the experimentally determined amplitude $A$ [Fig.~\ref{fig:3}\textcolor{blue}{(b)}], confirming the reduction of $A$ both for positive and negative applied fields, as a consequence of the variation of the spin-diffusion length for ${E_x<0}$ and of the transit time of spin-polarized electrons below the detection pad for ${E_x>0}$. The main discrepancy between experimental data and 2D simulations is that the latter underestimate the decrease of $A$ for ${E_x>0}$. We ascribe this deviation to the presence of a $z$ component of the electric field (not accounted for in the simulations), which modifies the built-in electric field of the Pt/Ge junction and thus modulates the transmission efficiency of electrons through the Schottky barrier \cite{Zucchetti2019a}. We stress, however, that 2D simulations capture the fundamental physical mechanisms underpinning spin-transport.

In conclusion, we have directly imaged the transport of spin-polarized electrons in Ge in either the diffusive or drift-diffusive regime. We experimentally demonstrate that, thanks to the finite spin lifetime, an electric field can strongly modulate the distance traveled by spin carriers before they depolarize. In this way, we are able to modulate by about one order of magnitude the spin-polarized electrons reaching the detector. The larger spin-transport length we measure is $40~\text{\textmu m}$, which in a purely diffusive regime can only be attained at cryogenic temperatures. Model predictions nicely match the experimental findings. Our results open the route to the exploitation of electric fields for guiding and controlling spins across spintronic devices.


\end{sloppypar}

%


\end{document}